\documentclass[aps,prd,preprint,eqsecnum,tightenlines,floats,
groupedaddress,showpacs]{revtex4}
\usepackage{amssymb} 
\usepackage{graphicx}
\usepackage{subfigure}
\usepackage{dcolumn}

\begin{document}

\date{February 10, 2022}

\title{Resonant excitation of the axion field during\\ 
the QCD phase transition}

\author{Pierre Sikivie and Wei Xue}
\affiliation{Department of Physics, University of Florida,
Gainesville, FL 32611, USA}

\vskip 1cm

\begin{abstract}

We find that the adiabatic fluctuations produced in the 
primordial plasma by cosmological inflation resonantly 
excite the axion field during the QCD phase transition 
by pumping axions from low momentum modes to modes with 
momentum up to of order $\sqrt{3} m$ where $m$ is the 
axion mass.  We derive the momentum distribution of the 
excited axions. The fraction of cold axions that get
excited is of order one if the axion mass is larger
than a few $\mu$eV.  The effect occurs whether inflation 
happens before or after the Peccei-Quinn phase transition.  

\end{abstract}

\maketitle

\section{Introduction}

The axion is a hypothetical particle postulated to 
solve the Strong CP Problem, i.e. to explain why the 
strong interactions conserve the discrete symmetries 
P and CP in spite of the fact that these symmetries 
are broken in the Standard Model of particle physics 
as a whole \cite{PQ,WW,inv}.  Axions are independently 
motivated by cosmology because a population of cold 
axions is produced during the QCD phase transition, 
when the temperature is of order 1 GeV ($k_B = \hbar 
= c = 1$) \cite{axdm}. In the simplest scenarios, 
axions account for the cold dark matter of the 
Universe if the axion mass is of order $10^{-5}$ 
eV, with large uncertainties \cite{axrev}.

The axion \cite{WW} is the quasi-Nambu-Golstone boson 
resulting from the spontaneous breaking of the U$_{\rm PQ}(1)$ 
anomalous global symmetry that Peccei and Quinn \cite{PQ} 
postulated.  The properties of the axion depend for the 
most part on a single parameter $f_a$ called the axion 
decay constant.  In particular the axion mass is 
\begin{equation}
m \simeq 6~\mu{\rm eV}
\left({10^{12}~{\rm GeV} \over f_a}\right)
\label{axmass}
\end{equation} 
at temperatures below 100 MeV.  At temperatures $T$ of 
order 1 GeV and higher, the axion mass is estimated to 
be \cite{axdm} 
\begin{equation}
m(T) \simeq 4 \cdot 10^{-9}~{\rm eV}
\left({10^{12}~{\rm GeV} \over f_a}\right)
\left({{\rm GeV} \over T}\right)^4
\label{highT}
\end{equation}
using the dilute instanton gas approximation \cite{GPY}.  
At intermediate temperatures, between 1 GeV and 100 MeV, 
the axion mass may be estimated from QCD lattice 
simulations \cite{Bors,Bonati}.  The precise manner
in which the axion mass turns on during the QCD
phase transition affects the quantitative results 
derived in the present paper.

An important source of uncertainty when estimating the 
abundance of cold axions, is whether the U$_{\rm PQ}$(1)
symmetry is spontaneously broken before or after the epoch 
of inflationary expansion \cite{infl,infl2} that homogenizes 
the very early universe.  U$_{\rm PQ}$(1) becomes spontaneously 
broken when the temperature falls below some critical value of 
order $f_a$, during the so-called PQ phase transition. If this 
phase transition occurs before inflation, the axion field gets 
homogenized by the inflationary expansion, so that at the start 
of the QCD phase transition the axion field oscillations begin 
with almost the same amplitude everywhere. Unless this initial 
amplitude happens to be zero, which would be an unexpected 
accident, axions are produced in a process of ``vacuum 
realignment" \cite{axdm}.  If the PQ phase transition 
occurs after inflation, the axion field enters the QCD 
phase transition with random initial amplitudes, 
uncorrelated from one QCD horizon volume to the 
next.  In that case cold axions are produced by 
vacuum realignment and by the decay of axion 
strings and domain walls \cite{axrev}.

In this paper we are interested in the momentum 
spectrum of cold axions at the end of the QCD
phase transition.  If inflation occurs after 
the PQ phase transition the axion field is 
homogenized by the inflationary expansion.   
However, as is well known, the axion fluid 
has perturbations which originate as quantum 
mechanical fluctuations in the axion field during 
inflation \cite{isoc}.  These perturbations have 
been studied extensively because they produce 
isocurvature density perturbations that may be 
inconsistent with the observed cosmic microwave 
background (CMB) anisotropies.  The observed 
anisotropies are consistent with purely adiabatic 
density perturbations.  The absence of isocurvature 
perturbations qualitatively requires 
$H_I \lesssim 10^{-5} f_a$ where $H_I$ is the Hubble 
constant during inflation.   

We are interested in length scales much smaller than 
those constrained by the CMB observations since the QCD 
horizon stretched to the present epoch is approximately 
$10^{17}$ cm, much smaller than the length scales probed 
in the CMB. On the other hand, since the power spectrum 
of density perturbations predicted by inflation is very 
nearly flat, observations or theoretical considerations 
at one scale are potentially relevant to the other scales 
as well.

The quantum mechanical fluctuations in the axion field
during inflation are not the only source of perturbations
in the axion fluid in case inflation occurs after the 
PQ phase transition.  A second source is the effect on the
axion fluid of the adiabatic perturbations in the primordial 
plasma \cite{adiab}.  For the wavevectors of interest to us, 
this second source of axion fluid perturbations dominates.  
Indeed we find that the adiabatic perturbations in the 
primordial plasma resonantly excite the axion field.  As 
far as we know, this effect has not been anticipated or 
commented on until now.  

The resonant excitation can be thought of as the process
$\tilde{\gamma}(\vec{k}) + \phi(\vec{0}) \rightarrow 
\phi(\vec{k})$, where $\tilde{\gamma}(\vec{k})$ is a 
plasmon and $\phi(\vec{k})$ an axion, both with momentum
$\vec{k}$.  It is shown in Section IV that for every 
comoving wavevector $\vec{k}$ there is a time interval 
during which this process conserves energy exactly or 
sufficiently well that it occurs efficiently.  During 
that time, axions are transferred from the very low 
momentum modes that they occupy initially to modes 
of momentum up to of order $\sqrt{3} m$. The process 
transfers energy from the plasma oscillations to the 
axion fluid.  The process is qualitatively the same 
whether inflation occurs before or after the PQ phase 
transition because in either case the initial momenta 
of the axions are much less than the transferred momenta.  
However, it is possible to derive the momentum spectrum
of cold axions with the greatest precision in case 
inflation occurs after the PQ phase transition, and 
this is one of our goals here.  Our treatment is 
complete only for modes that enter the horizon well 
before the QCD phase transition.  The analysis of 
modes that enter the horizon near the start of and 
during the QCD phase transition is left for future 
work.

The outline of our paper is as follows. In Section II, 
we state the equation of motion for the axion field,
the spectrum of initial perturbations in the axion 
field, and the spectrum of adiabatic perturbations 
in the primordial plasma.  In Section III we evolve 
the initial perturbations in the axion field and 
derive the resulting axion momentum spectrum.  In 
Section IV we derive the perturbations in the axion 
field induced by the adiabatic fluctuations in the 
primoridial plasma and the resulting axion momentum 
spectrum.  Section V summarizes our results.

\section{Equation of motion and initial conditions}

Ignoring $\lambda \phi^4$ self-interactions, the axion 
field $\phi(\vec{x},t)$ satisfies a linear equation of 
motion: 
\begin{equation}
D^\mu \partial_\mu \phi(\vec{x},t) 
- m^2(T(\vec{x},t)) \phi(\vec{x},t) = 0
\label{eom}
\end{equation}
where $m(T)$ is the temperature-dependent axion mass.
We are interested in scalar perturbations only.
So we may adopt the so-called conformal-Newtonian
gauge for the space-time metric:
\begin{equation}
g_{\alpha\beta}(x) dx^\alpha dx^\beta = 
- (1 + 2 \Psi(\vec{x},t)) dt^2 
+ a^2(t)(1 + 2 \Phi(\vec{x},t)) d\vec{x}\cdot d\vec{x}~~\ .
\label{metric}
\end{equation}
During the QCD phase transition, ignoring small corrections
due to changes in the number of thermal degrees of freedom
as a function of temperature, the scale factor 
$a(t) \propto \sqrt{t}$ and the Hubble expansion rate 
$H(t) = {1 \over 2t}$.  When only zeroth and first 
order terms in $\Psi(\vec{x},t)$, $\Phi(\vec{x},t)$ 
and the temperature fluctuations $\delta T(\vec{x},t)$
are included, Eq.(\ref{eom}) becomes
\begin{eqnarray}
&~&(- 1 + 2 \Psi) \partial_t \partial_t \phi 
+ \partial_t \Psi \partial_t \phi 
+ {1 \over a^2} \partial_j \Psi \partial_j \phi
\nonumber\\
&+& (1 - 2 \Phi) ({1 \over a^2} \partial_j \partial_j \phi
- 3 H \partial_t \phi) - 3 (\partial_t \Phi + 2 H \Phi - 2 H \Psi) 
\partial_t \phi
\nonumber\\
&+& {1 \over a^2} \partial_j \Phi \partial_j \phi 
- [m^2(T(t)) + {d m^2 \over d T} \delta T] \phi = 0
\label{eom2}
\end{eqnarray}
where $j = 1, 2, 3$ labels the space directions.

At temperatures of order 1 GeV, the axion mass is given 
by Eq.~(\ref{highT}).  It implies the time dependence:
\begin{equation}
m(t) \simeq 0.9 \cdot 10^6 {1 \over {\rm sec}}
\left({t \over 10^{-7}~ {\rm sec}}\right)^2
\left({10^{12}~{\rm GeV} \over f_a}\right)~~\ .
\label{axmass3}
\end{equation}
We define $t_1$ such that $m(t_1) t_1 = 1$.  
Eq.~(\ref{axmass3}) implies 
\begin{equation}
t_1 \simeq 2 \cdot 10^{-7}~{\rm sec} 
\left({f_a \over 10^{12}~{\rm GeV}}\right)^{1 \over 3}
\label{time1}
\end{equation}
at which time the temperature is
\begin{equation}
T_1 \simeq 1~{\rm GeV} 
\left({10^{12}~{\rm GeV} \over f_a}\right)^{1 \over 6}~~\ .
\label{Temp1}
\end{equation}
At lower temperatures, the axion mass may be obtained by QCD 
lattice simulations.  The results of Sz. Borsanyi et al. 
\cite{Bors} and C. Bonati et al. \cite{Bonati} are 
consistent with one another and with the temperature 
dependence given in Eq.~(\ref{highT}) down to 
$T_2 \simeq 160$ MeV, where $m(T)$ turns over 
and becomes its zero temperature value given in 
Eq.~(\ref{axmass}).  The age of the universe at 
temperature $T_2$ is  $2 \cdot 10^{-5}$ sec 
\cite{Husdal} when the thinning out of thermal 
degrees of freedom during the QCD phase transition 
is taken ito account.  In our estimates below, to 
simplify the calculations somewhat,  we will ignore 
the thinning out of thermal degrees of freedom and 
assume 
\begin{eqnarray}
m(t) &=& {1 \over t_1} \left({t \over t_1}\right)^2 
= m \left({t \over t_2}\right)^2~~~~{\rm for}~t<t_2
\nonumber\\
&=& m 	~~~~~~~~~~~~~~~~~~~~~~~~~~~~~~~{\rm for}~t>t_2~~\ .
\label{mt}
\end{eqnarray}
This requires $t_2 = \sqrt{m t_1^3} \simeq 10^{-5}$ sec.
In principle, our approach can be used for any smooth turn-on 
of the axion mass.   It fails if the axion mass has a discontinuity 
as a function of temperature, as may occur if the QCD phase transition 
is first order.

\begin{figure}
\begin{center}
\includegraphics[height=100mm]{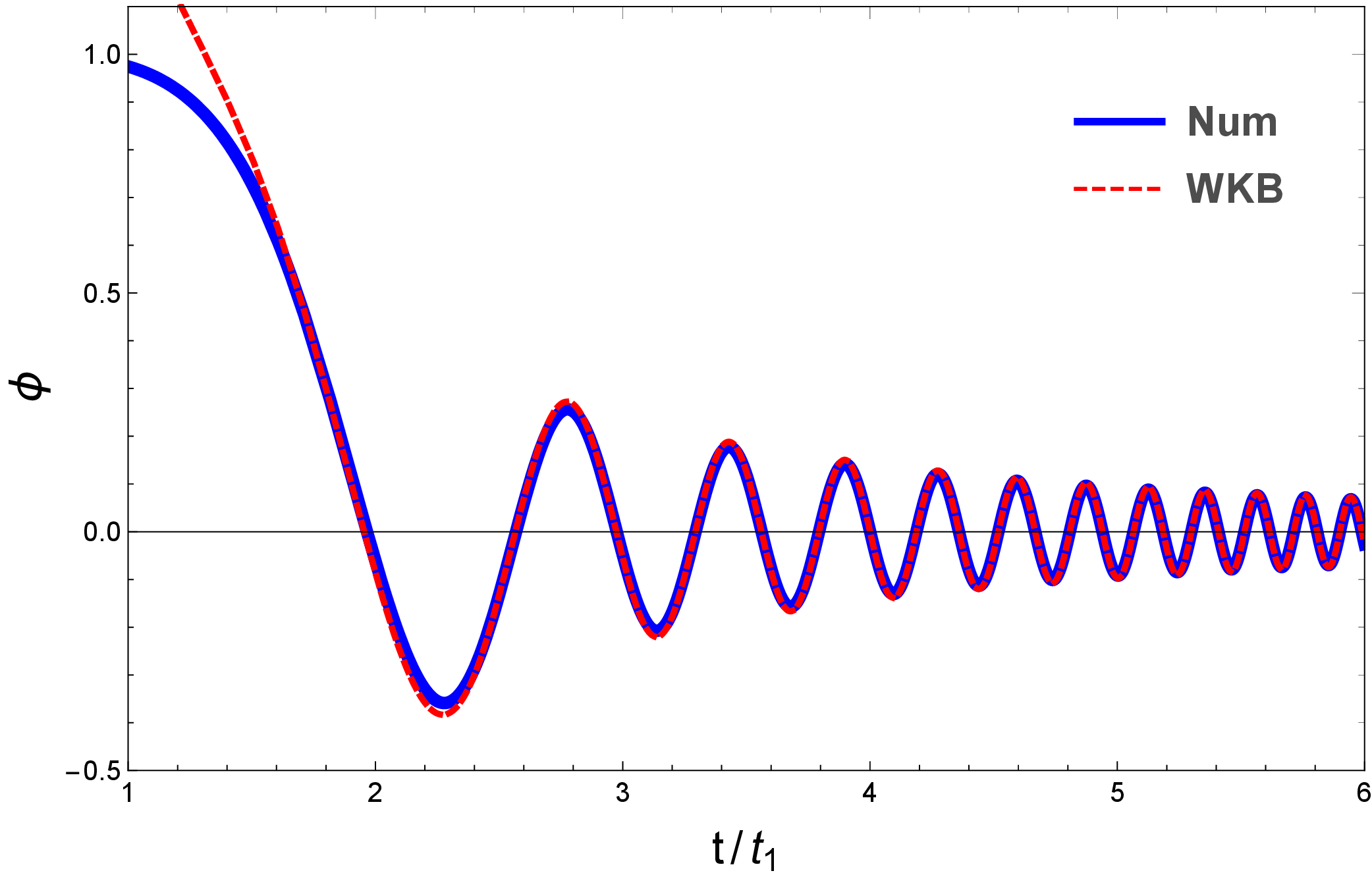}
\vspace{0.3in}
\caption{Plot of the axion field zero mode $\phi^{(0)}(t)$
as it begins to oscillate at the onset of the QCD phase
transition.   The continuous line is the result of
numerically integrating Eq.~(\ref{0th}) starting with
an initial constant value. The dashed line is the
time dependence given in Eq.~(\ref{0sol2}) for an
appropriate choice of the phase $\delta$ and
amplitude $B$.}
\end{center}
\label{fig1}
\end{figure}

The zeroth order equation for the axion field is
\begin{equation}
({d^2 \over dt^2} + {3 \over 2t} {d \over dt}
+ m^2(t))\phi^{(0)}(t) = 0~~\ .
\label{0th}
\end{equation}
For $t<<t_1$, when the axion mass is negligible, 
Eq.~(\ref{0th}) is solved by 
\begin{equation}
\phi^{(0)}(t) = A + D {1 \over \sqrt{t}}
\label{0sol}
\end{equation}
where $A$ and $D$ are constants.  We set $D=0$ since
${d \over dt} \phi^{(0)} = 0$ at the end of inflation.
For $t>>t_1$, the change in the axion mass is adiabatic 
(${dm \over dt} << m^2$) so that 
\begin{equation}
\phi^{(0)}(t) = {B \over \sqrt{t_1 m(t)}}
\left({t_1 \over t}\right)^{3 \over 4}
\cos\left(\int_0^t dt^\prime m(t^\prime) + \delta\right)~~\ .
\label{0sol2}
\end{equation}
The amplitude $B$ and phase $\delta$ can be obtained 
in terms of $A$ by solving Eq.~(\ref{0th}) numerically.
Fig. 1 shows the function $\phi^{(0)}(t)$ for $A=1$.
After the QCD phase transition, the energy density in the 
zero mode $\phi^{(0)}$ is 
\begin{equation}
\rho^{(0)}(t) = {B^2 \over 2 t_1} 
\left({t_1 \over t}\right)^{3 \over 2} m
\label{enden0}
\end{equation}
which is indeed the standard result \cite{axdm}.

The first order equation for the perturbations 
$\phi^{(1)}(\vec{x},t) = \phi(\vec{x},t) - \phi^{(0)}(t)$
in the axion field is
\begin{equation}
\partial_t^2 \phi^{(1)}(\vec{k},t) + {3 \over 2t} \phi^{(1)}(\vec{k},t)
+ ({\vec{k}\cdot\vec{k} \over a^2(t)} + m^2(t)) \phi^{(1)}(\vec{k},t)
= j(\vec{k},t)
\label{1st}
\end{equation}
where 
\begin{eqnarray}
j(\vec{k},t) &=& 2 \Psi(\vec{k},t) \partial_t^2 \phi^{(0)}(t)
\nonumber\\
&+& (\partial_t \Psi(\vec{k},t) - 3 \partial_t \Phi(\vec{k},t) 
+ 6 H(t) \Psi(\vec{k},t)) \partial_t \phi^{(0)}(t)
\nonumber\\
&-& {d m^2 \over d T}(T(t)) \delta T(\vec{k},t) \phi^{(0)}(t)~~\ .
\label{source}
\end{eqnarray}
The $\phi^{(1)}(\vec{k},t)$ are the Fourier modes of 
$\phi^{(1)}(\vec{x},t)$, defined by
\begin{equation}
\phi^{(1)}(\vec{x},t) = \int {d^3 k \over (2 \pi)^3} 
\phi^{(1)}(\vec{k},t) e^{i \vec{k}\cdot\vec{x}}~~\ .
\label{Four}
\end{equation}
$\Psi(\vec{k},t)$, $\Phi(\vec{k},t)$ and $\delta T(\vec{k},t)$ 
are likewise the Fourier modes of $\Psi(\vec{x},t)$, $\Phi(\vec{x},t)$
and $\delta T(\vec{x},t)$.  

The quantum mechanical fluctuations in the inflaton field  
produce adiabatic fluctuations.  Provided anisotropic stresses 
in the primordial plasma are small, as is normally the case, 
$\Psi(\vec{k},t) = - \Phi(\vec{k},t)$ \cite{Dodelson}.  We 
will assume this to be true henceforth.  The power spectrum 
of $\Phi$ fluctuations when they reenter the horizon 
(k = a(t) H(t)) is  
\begin{equation}
<\Phi(\vec{k},t) \Phi^*(\vec{k}^\prime,t)>|_{k = a H} 
\equiv <\Phi_p(\vec{k}) \Phi_p^*(\vec{k}^\prime)> = 
(2 \pi)^3 \delta(\vec{k} - \vec{k}^\prime)
{8 \pi G \over 9 k^3} {H_I^2 \over \epsilon}
\label{adpow}
\end{equation}
where $G$ is Newton's gravitational constant, $H_I$ is 
the expansion rate during inflation, and $\epsilon \equiv
{d \over dt}{1 \over H_I(t)}$.  The amplitude of adiabatic 
perturbations is derived from observations of the cosmic 
microwave background anisotropies on scales of order the 
present horizon:
\begin{equation}
{8 \pi G \over 9 k^3} {H_I^2 \over \epsilon} = 
{50 \pi^2 \over 9 k^3} \left({k \over H_0}\right)^{n-1}
\delta_H^2 \left({\Omega_m \over D_1(t_0)}\right)^2
\label{amp}
\end{equation}
where $H_0$ is the present expansion rate, $n \simeq 0.96$ 
is the scalar power index, $\delta_H \simeq 4.6 \cdot 10^{-5}$
is the amplitude of adiabatic density perturbations when they 
reenter the horizon, $\Omega_m \simeq$ 0.3 is the fraction of 
the critical energy density in matter today, and 
$D_1(t_0) \simeq 0.8$ is the growth function. Eq.~(\ref{amp}) 
and the various parameter estimates are taken from ref. 
\cite{Dodelson}.  Eq.~(\ref{amp}) applies on wavevector scales 
of order $H_0^{-1}$.  We extrapolate it, over approximately 13 
orders of magnitude, to modes that enter the horizon shortly 
before the QCD phase transition. This yields
\begin{equation}
{8 \pi G H_I^2 \over 9 \epsilon} \equiv C \simeq 4.9 \cdot 10^{-9}~~\ .
\label{adpow2}
\end{equation}
Our results in Section IV are expressed in terms of $C$, with 
the value given in Eq.~(\ref{adpow2}) substituted only at the
end to make final estimates, in case the assumptions made above 
about the inflationary model need to be modified.

After they have entered the horizon, the $\Phi$ 
fluctuations oscillate as follows \cite{Dodelson}
\begin{eqnarray}
\Phi(\vec{k},t) &=& 3 \Phi_p(\vec{k})  
\left({\sqrt{3} \over 2 k \sqrt{t t_1}}\right)^3
\left[\sin \left(2 k \sqrt{t t_1 \over 3}\right)
- {2 k \sqrt{t t_1 \over 3}} 
\cos \left(2 k \sqrt{t t_1 \over 3}\right)\right]
\nonumber\\
&=& - {9 \over 4 k^2 t t_1}\Phi_p(\vec{k})
\cos \left(2 k \sqrt{t t_1 \over 3}\right)~~~~~{\rm for}
~~k \sqrt{t t_1} >> 1  ~~\ .
\label{timedep}
\end{eqnarray}
We have normalized the scale factor so that $a(t_1) = 1$.  
$\vec{k}$ is thus the physical wavevector at time $t_1$.  
The $\Phi$ fluctuations are accompanied by fluctuations in 
the temperature \cite{Dodelson}
\begin{eqnarray}
{\delta T(\vec{k},t) \over T(t)}
&=& ({1 \over 2} + {2 k^2 t t_1 \over 3}) \Phi(\vec{k},t) +
{t \over 4} {d \over dt} \Phi(\vec{k},t)
\nonumber\\
&=& - {3 \over 2} \Phi_p(\vec{k}) 
\cos \left({2 k \sqrt{t t_1 \over 3}}\right)~~~{\rm for}
~~~~~ k \sqrt{t t_1} >> 1~~ \ . 
\label{tempfl}
\end{eqnarray}
The quantum mechanical fluctuations in the axion field, 
generated during inflation, have power spectrum
\begin{equation}
<\delta\phi(\vec{k},t) \delta\phi^*(\vec{k}^\prime,t)>|_{k = aH}
\equiv <\phi_p(\vec{k}) \phi_p^*(\vec{k}^\prime)>
= (2 \pi)^3 \delta(\vec{k} - \vec{k}^\prime) {H_I^2 \over 2 k^3}
\label{axfl}
\end{equation}
when they reenter the horizon.  

There are therefore two kinds of perturbations in the axion field
\begin{equation}
\phi(\vec{x},t) = \phi^{(0)}(t) + \phi^{(1)}_{\delta\phi}(\vec{x},t)
+ \phi^{(1)}_\Phi(\vec{x},t)
\label{axpert}
\end{equation}
where the $\phi^{(1)}_{\delta\phi}$ perturbations originate as 
quantum mechanical fluctuations in the axion field during inflation 
and the $\phi^{(1)}_\Phi$ perturbations are induced in the axion 
field by the adiabatic fluctuations in the primordial plasma.  
The two kinds of perturbations are assumed to be uncorrelated, i.e. 
\begin{equation}  
<\Phi_p(\vec{k}) \phi_p(\vec{k}^\prime)> = 0~~\ .
\label{uncor}
\end{equation}
The fluid density in non-zero mode axions is then a simple sum:
\begin{equation}
\rho^{(1)} = \rho^{(1)}_{\delta\phi} + \rho^{(1)}_\Phi
\label{pert}
\end{equation}
where $\rho^{(1)}_{\delta\phi}$ and $\rho^{(1)}_\Phi$ are the 
contributions from $\phi^{(1)}_{\delta\phi}$ and $\phi^{(1)}_\Phi$ 
respectively. They are evaluated in Sections III and IV respectively.

\vskip 1cm

\section{Quantum-mechanical fluctuations of the axion field}

In this section we set $j(\vec{x},t) = 0$ and derive how 
the axion field perturbations evolve in the absence of 
adiabatic perturbations in the primordial plasma.  The 
equation of motion is 
\begin{equation}
[\partial_t^2 + {3 \over 2t} \partial_t 
+ k^2 {t_1 \over t} + m^2(T(t))] \phi^{(1)}(\vec{k},t) = 0~~\ .
\label{eom0}
\end{equation}
It may also be written
\begin{equation}
(\partial_t^2 + \Omega^2(k,t)) B^{(1)}(\vec{k},t) = 0
\label{eomb}
\end{equation}
where $B^{(1)}(\vec{k},t)$ is defined by 
\begin{equation}
\phi^{(1)}(\vec{k},t) = \left({t_1 \over t}\right)^{3 \over 4}
B^{(1)}(\vec{k},t)
\label{defb}
\end{equation}
and 
\begin{equation}
\Omega(k,t) = \sqrt{{3 \over 16 t^2} + k^2 {t_1 \over t} + m^2(t)}~~\ .
\label{Bom}
\end{equation}
For $t << t_1$, when the axion mass is negligible, Eq.~(\ref{eom0})
is solved exactly by
\begin{equation}
\phi^{(1)}(\vec{k},t) = \phi_p(\vec{k}) {1 \over 2k \sqrt{t t_1}}
\sin(2 k \sqrt{t t_1})~~ \ ,
\label{exact} 
\end{equation}
$\phi_p(\vec{k})$ being the initial value of the perturbation.

\begin{figure}
\begin{center}
\includegraphics[height=100mm]{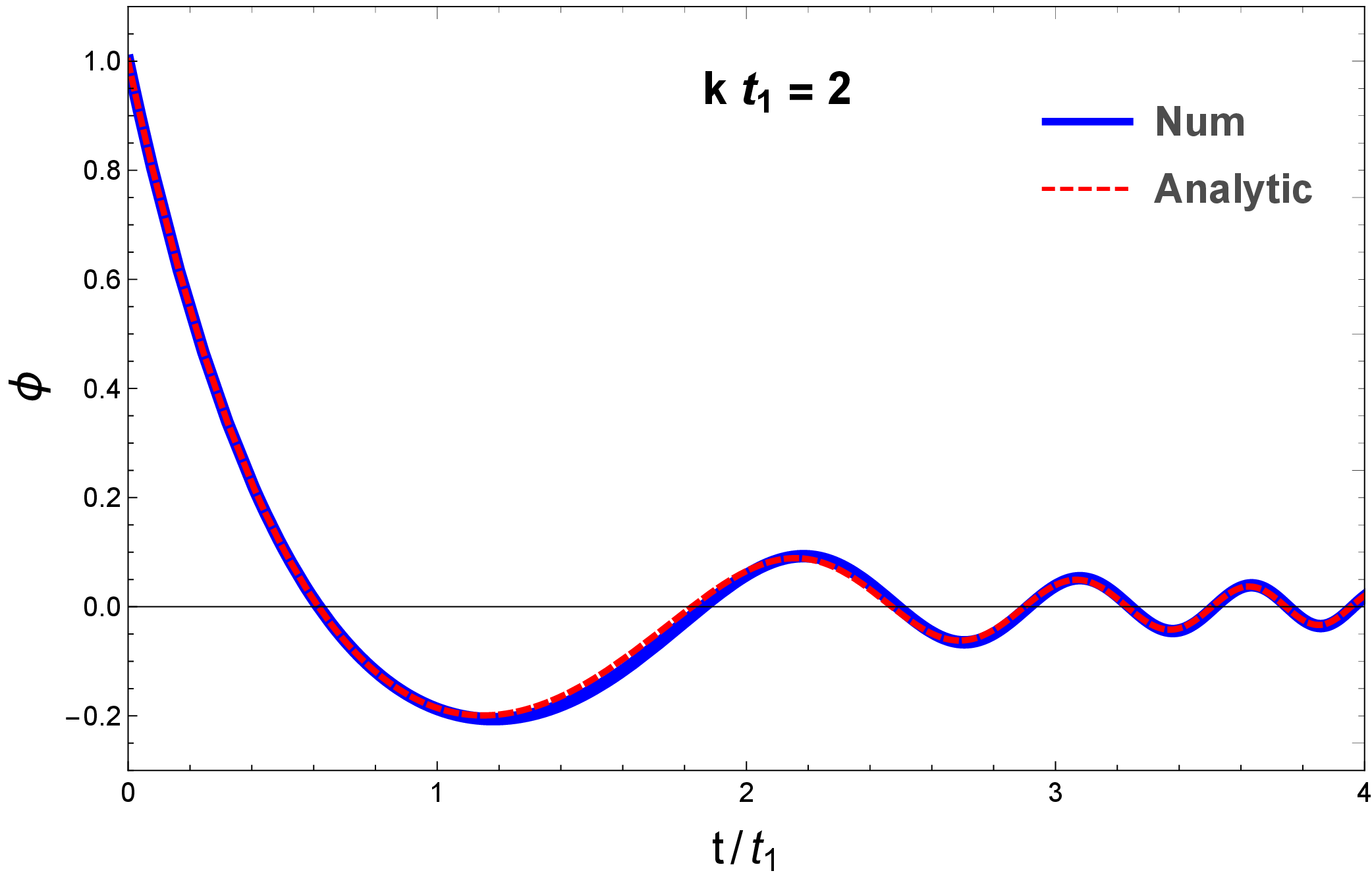}
\caption{Plot of $\phi^{(1)}(\vec{k},t)$ for $k = 2/t_1$.
The continuous line is the result of numerically solving
Eq.~(\ref{eom0}).  The dashed line is the time dependence   
given in Eq.~(\ref{allt}).}
\end{center}
\label{fig2}
\end{figure}

For $k >> {1 \over t_1}$ the turn on of the axion mass is 
adiabatic and therefore
\begin{equation}
\phi^{(1)}(\vec{k},t) = {C(\vec{k}) \over \sqrt{\omega(k,t)}}
\left({t_1 \over t}\right)^{3 \over 4}
\sin \left( \int_0^t dt^\prime \omega(k,t^\prime)\right)
\label{adsol}
\end{equation}
solves Eq.~(\ref{eom0}) for $t \gtrsim t_1$ with 
\begin{equation}
\omega(k,t) = \sqrt{k^2 {t_1 \over t} + m^2(t)}~~\ .
\label{lom}
\end{equation}
The RHS of Eq.~(\ref{exact}) equals that of Eq.~(\ref{adsol}) for 
$t << t_1$ provided $C(\vec{k}) = \phi_p(\vec{k})/(2 \sqrt{k} t_1)$.
We have therefore 
\begin{equation}
\phi^{(1)}(\vec{k},t) = {\phi_p(\vec{k}) \over 2 t_1 \sqrt{k \omega(k,t)}}
\left({t_1 \over t}\right)^{3 \over 4} 
\sin \left( \int_0^t dt^\prime \omega(k,t^\prime)\right) 
\label{allt}
\end{equation}
for all $t$ when $j(\vec{k},t) = 0$ and $k >> {1 \over t_1}$.  
Eq.~(\ref{allt}) was compared with the result of numerically 
solving Eq.~(\ref{eom0}), and found to be practically 
indistinguishable from it for $k \geq 2/t_1$.   Fig. 2 
shows the comparison for $k = 2/t_1$. 

Eq.~(\ref{allt}) implies for $t >> t_1$
\begin{eqnarray}
\rho^{(1)}_{\delta\phi} &=& \left({t_1 \over t}\right)^{3 \over 2}
\int {d^3 k \over (2 \pi)^3} \int {d^3 k^\prime \over (2 \pi)^3} 
<\phi_p(\vec{k}) \phi_p(\vec{k}^\prime)> {m \over 8 t_1^2 k} 
\nonumber\\
&\simeq& m \left({t_1 \over t}\right)^{3 \over 2} {H_I^2 \over 16 t_1^2}
\int_{1 \over t_1} {d^3 k \over (2 \pi)^3} {1 \over k^4}
\nonumber\\
&\simeq& m \left({t_1 \over t}\right)^{3 \over 2}
\left({H_I \over 2 \pi}\right)^2 {1 \over 8 t_1}~~\ ,
\label{phi1}
\end{eqnarray}
including the contribution of all modes that enter the horizon 
before $t_1$.  $\rho^{(1)}_{\delta\phi}$ is very small compared 
to $\rho^{(0)}$ since the observed absence of isocurvature 
perturbations implies $H_I \lesssim 10^{-5} f_a$. 

\vskip 1cm 

\section{Resonant excitation of the axion field}

We now turn to the effects that result from the adiabatic 
fluctuations in the plasma.  Fig. 3 summarizes our conclusions.
It shows the spectrum of excited axions.  The horizontal axis 
is divided into three regions.  The region $k_* < k < k_c$ is 
dominated by axions which are resonantly excited at a time 
intermediate between $t_1$ and $t_2$.  The region $k < k_*$ 
is dominated by axions which are excited by a near resonance 
effect which occurs from the time they are resonantly 
excited till time $t_2$.  The axions that are resonantly 
excited after time $t_2$ have wavevectors $k > k_c$.  

\subsection{General remarks}

Eq.~(\ref{1st}) shows that $j(\vec{k},t)$ acts as a 
source for $\phi^{(1)}(\vec{k},t)$.  This source is 
turned off for $t << t_1$, because $\partial_t \phi^{(0)}$ 
and ${\partial m^2 \over \partial T}$ are negligible then.  
We rewrite Eqs.~(\ref{1st}) and (\ref{source}) in terms of 
$B^{(0)}(t)$ defined by 
\begin{equation}
\phi^{(0)}(t) = B^{(0)}(t) \left({t_1 \over t}\right)^{3 \over 4}
\label{defin}
\end{equation}
and $B^{(1)}(\vec{k},t)$ defined by Eq.~(\ref{defb}),
and use Eq.~(\ref{0th}) to eliminate $\partial_t^2 \phi^{(0)}$.
This yields
\begin{eqnarray}
(\partial_t^2 + \Omega^2(k,t))B^{(1)}(\vec{k},t) &=&
- 4 \partial_t \Phi(\vec{k},t) \partial_t B^{(0)}(t) 
\nonumber\\
+ [- {d  m^2 \over d T} \delta T(\vec{k},t)
&+& 2 m^2(t) \Phi(\vec{k},t) 
+ 6 H(t) \partial_t \Phi(\vec{k},t)]B^{(0)}(t)~~\ .
\label{Beom}
\end{eqnarray}
We substitute the late time ($k \sqrt{t t_1} >>1$) behaviors 
of $\Phi(\vec{k},t)$ and $\delta T (\vec{k},t)$, given in 
Eqs.~(\ref{timedep}) and (\ref{tempfl}), and that of 
$B^{(0)}(t)$ implied by Eq.~(\ref{0sol2}).  This yields
\begin{equation}
(\partial_t^2 + \omega^2(k,t)) B^{(1)}(\vec{k},t) = f(\vec{k},t)
\label{dho}
\end{equation}
with
\begin{eqnarray}
f(\vec{k},t) &=&  {\Phi_p(\vec{k}) B \over \sqrt{t_1 m(t)}}
[\left({3 \over 2} {d m^2 \over d T}(t) T(t) 
- {9 m^2(t) \over 2 k^2 t t_1}\right) 
\cos \left(\int_0^t dt^\prime m(t^\prime) + \delta\right)
\cos\left(2 k \sqrt{t t_1 \over 3}\right) 
\nonumber\\
&+& {9 m(t) \over k t \sqrt{3 t t_1}}
\sin \left(\int_0^t dt^\prime m(t^\prime) + \delta\right)
\sin \left(2 k \sqrt{t t_1 \over 3}\right)]
\label{force}
\end{eqnarray}
for $k\sqrt{t t_1} >> 1$. 

\begin{figure}
\begin{center}
\includegraphics[height=100mm]{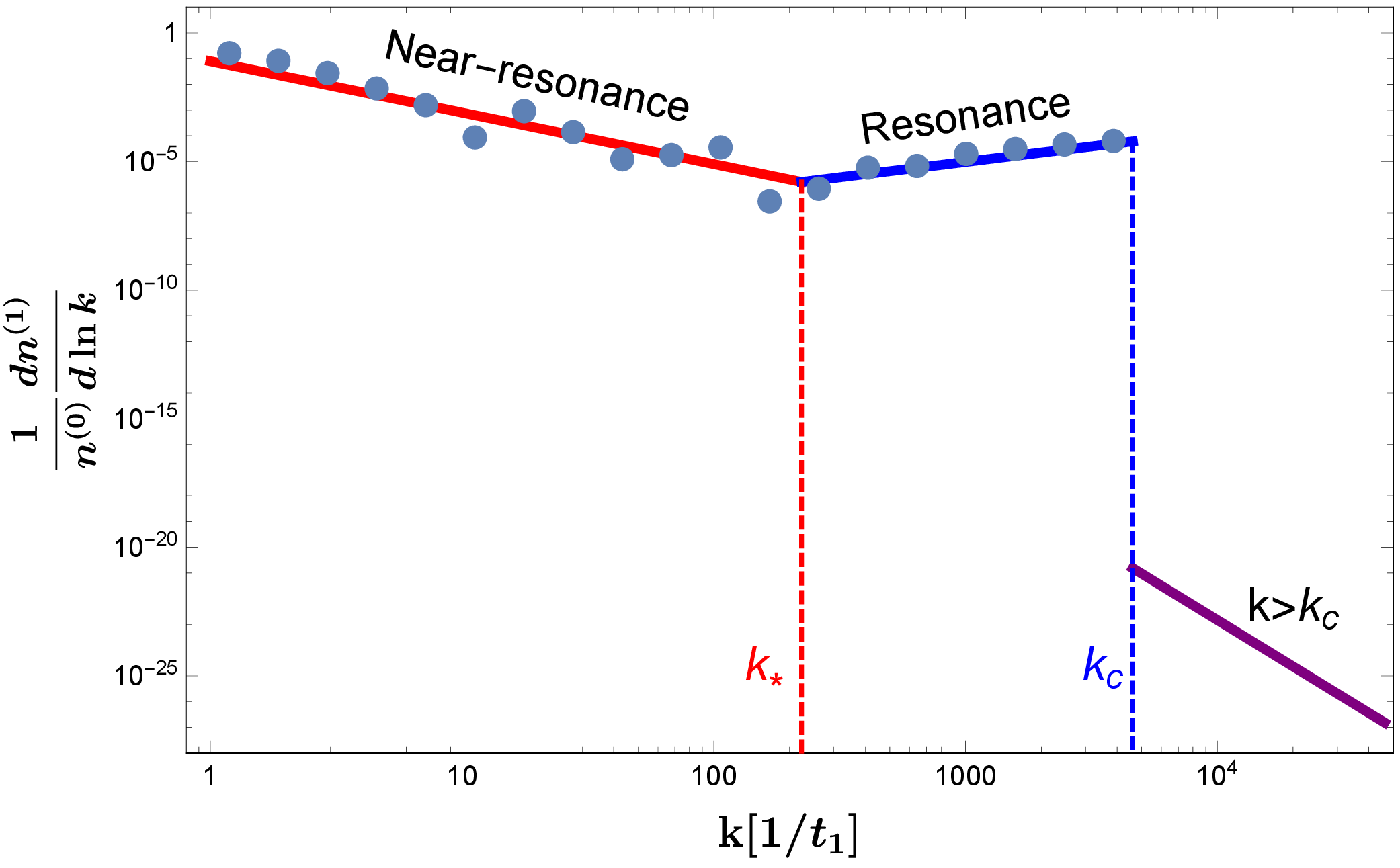}
\caption{Spectrum of axions excited by the resonance proper 
(blue and purple lines) and near resonance effects (red line), 
for $m = 1 \mu$eV.  Only the dominant contribution is shown 
for given $k$.  The contribution to the axion spectrum from 
the quantum mechanical fluctuations in the axion field, 
discussed in Section III, is not shown here.  The grey dots 
indicate the results from numerically solving Eq.~(\ref{dho}).}
\end{center}
\label{fig3}
\end{figure}

\begin{figure}
\begin{center}
\includegraphics[height=100mm]{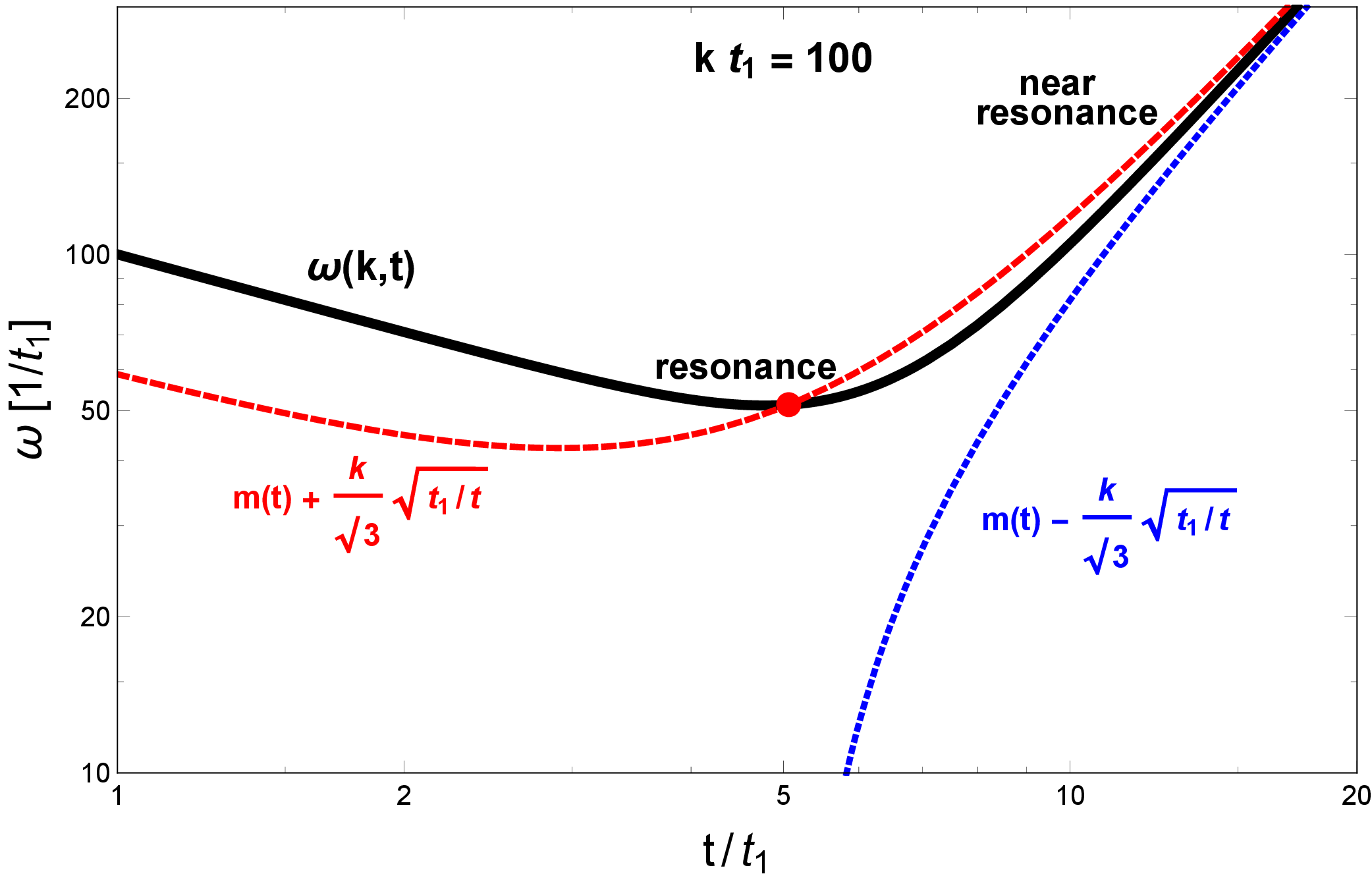}
\caption{Plot of the frequencies $\omega(k,t)$ and
$\omega_\pm(k,t) = m(t) \pm {k \over \sqrt{3}} \sqrt{t_1 \over t}$
as a function of time for $k = 100/t_1$.}
\end{center}
\label{fig4}
\end{figure}

Resonant excitation of the axion field occurs because 
$f(\vec{k},t)$ includes terms that oscillate briefly with 
frequency $\omega(k,t)$.  Indeed the frequencies that appear
in $f(\vec{k},t)$ are $\pm~ m(t) \pm ~{k \over \sqrt{3}}
\sqrt{t_1 \over t}$. Resonance occurs when 
\begin{equation}
\omega(k,t) = \sqrt{m^2(t) + k^2 {t_1 \over t}} = 
m(t) + {k \over \sqrt{3}} \sqrt{t_1 \over t}~~\ ,
\label{reson}
\end{equation}
i.e. at time $t_k$ such that
\begin{equation}
m(t_k) = {k \over \sqrt{3}} \sqrt{t_1 \over t_k}~~\ .
\label{tk}
\end{equation}
Henceforth we assume a specific time dependence for the 
axion mass:  Eq.~(\ref{axmass3}) for $t<t_2$ and
Eq.~(\ref{axmass}) for $t>t_2$.  Let us call $k_c$
the comoving wavevector magnitude of modes that are 
resonantly excited at time $t_2$.  Eq.~(\ref{tk}) 
with $t_k=t_2$ and $m(t_k) = m$ implies
\begin{equation}
k_c =  m \sqrt{3 t_2 \over t_1}~~\ .
\label{kc}
\end{equation}
Modes with comoving wavevector magnitude $k_c$ have 
physical wavevector magnitude equal to $\sqrt{3} m$
at time $t_2$.  Modes with $k > k_c$ are resonantly 
excited at time
\begin{equation}
t_k = \left({k \over k_c}\right)^2 t_2~~\ , 
\label{tkl}
\end{equation}
which is after $t_2$.  Those modes are discussed 
in Subsection IV.D.  Modes with $k < k_c$ are resonantly 
excited at time
\begin{equation}
t_k = \left({k \over k_c}\right)^{2 \over 5} t_2
\label{tke}
\end{equation}
which is before $t_2$, when the axion mass is still turning 
on.  (Note that the larger its comoving wavevector magnitude 
$k$, the earlier a mode enters the horizon, but the later it 
gets resonantly excited.)  The frequencies $\omega(k,t)$ and 
$\omega_\pm(k,t) \equiv 
m(t) \pm {k \over \sqrt{3}} \sqrt{t_1 \over t}$ are plotted 
in Fig. 4 for $k = 100/t_1$.  The figure shows that, after 
they are equal at time $t_k$, $\omega(k,t)$ and $\omega_+(k,t)$ 
stay close to each other. $\omega_-(k,t)$ soon comes close 
$\omega(k,t)$ as well. A near resonance excitation of the 
axion field occurs from approximately time $t_k$ till $t_2$, 
as discussed in Subsection IV.C.  The resonance proper near  
time $t_k$ is discussed in Subsection IV.B.

We write 
\begin{equation}
B^{(1)}(\vec{k},t) = \Phi_p(\vec{k}) (a(k,t) {\cal B}(k,t) + c.c.)
\label{sep}
\end{equation}
where 
\begin{equation}
{\cal B}(k,t) = {1 \over \sqrt{\omega(k,t)}}
e^{i \int_0^t dt^\prime \omega(k,t^\prime)}
\label{calB}
\end{equation}
solves the equation of motion in the absence of source.  We 
substitute Eq.~(\ref{sep}) into Eq.~(\ref{dho}), neglect 
${d^2 a \over dt^2}$ relative to $\omega {d a \over dt}$, 
and keep only terms in $f(\vec{k},t)$ that produce resonance
or near resonance.  This yields
\begin{equation}
{d \over d t} a(k,t) = {B \over 2 i \sqrt{\omega(k,t)}}
[g_+(k,t) e^{i(h_+(k,t) + \delta)} + g_-(k,t) e^{i(h_-(k,t) + \delta)}]
\label{dadt}
\end{equation}
where 
\begin{equation}
h_\pm(k,t) = \int_0^t dt^\prime [\pm {k \over \sqrt{3}} \sqrt{t_1 \over 
t^\prime} + m(t^\prime) - \omega(k,t^\prime)]
\label{ht}
\end{equation}
and 
\begin{equation}
g_\pm(k,t) = {1 \over 4 \sqrt{t_1 m(t)}}
\left[{3 \over 2} {d m^2 \over d T}(t) T(t) 
- {9 m^2(t) \over 2 k^2 t_1 t}
\mp {9 m(t) \over k t \sqrt{3 t_1 t}} \right]~~\ .
\label{gkt}
\end{equation}
The initial value of $\Phi_p(\vec{k}) a(k,t)$ is 
${\phi_p(\vec{k}) \over 4it_1 \sqrt{k}}$ since the initial 
value of $\phi^{(1)}(\vec{k},t)$ is $\phi_p(\vec{k})$;
see Eq.~(\ref{allt}).  Its final value is therefore 
\begin{equation}
\Phi_p(\vec{k}) a(k) = {\phi_p(\vec{k}) \over 4 i t_1 \sqrt{k}} 
+ {\Phi_p(\vec{k}) B \over 2 i} (I_+(k) + I_-(k))
\label{Dak}
\end{equation}
with 
\begin{equation}
I_\pm(k) = \int_0^\infty dt {g_\pm(k,t) \over \sqrt{\omega(k,t)}}
e^{i(h_\pm(k,t) + \delta)}~~\ .
\label{Ike}
\end{equation}
Combining Eqs. (\ref{defb}), (\ref{sep}), (\ref{calB}) 
and (\ref{Dak}-\ref{Ike}), we obtain
\begin{eqnarray}
\phi^{(1)}(\vec{k},t) &=& {1 \over \sqrt{m}}
\left({t_1 \over t}\right)^{3 \over 4}
[{\phi_p(\vec{k}) \over 2 t_1 \sqrt{k}}
\sin(mt + \delta(k,t))
\nonumber\\
&+& |I(k)| \Phi_p(\vec{k}) B  \sin(mt + \delta^\prime(k,t))]
\label{finph}
\end{eqnarray}
for $t >> t_2$, where $I(k) = I_+(k) + I_-(k)$, 
\begin{equation}
\delta(k,t) = \int_0^t dt^\prime \sqrt{k^2 {t_1 \over t^\prime}
+ m^2(t^\prime)} - m t
\label{del}
\end{equation}
and
\begin{equation}
\delta^\prime(k,t) = \delta(k,t) + arg(I(k))~~\ .
\label{delp}
\end{equation}  
$\delta(k,t)$ and $\delta^\prime(k,t)$ vary only 
logarithmically with time after $t_2$.

Eqs.~(\ref{finph}), (\ref{adpow}), (\ref{adpow2}), 
(\ref{axfl}) and (\ref{uncor}) imply that the fluid 
density in non-zero mode axions is for $t>>t_2$
\begin{equation}   
\rho^{(1)} = m^2 <(\phi^{(1)}(\vec{x},t))^2>
= \rho^{(1)}_{\delta \phi}(t) + \rho^{(1)}_\Phi(t)
\label{1stden}
\end{equation}
with $\rho^{(1)}_{\delta \phi}$ given by Eq.~(\ref{phi1}) and
\begin{equation}
\rho^{(1)}_\Phi(t) = m \left({t_1 \over t}\right)^{3 \over 2}
{C B^2 \over 2} \int {d^3 k \over (2 \pi)^3} {1 \over k^3} |I(k)|^2~~\ . 
\label{phi2}
\end{equation}
We may rewrite this as
\begin{equation}
\rho^{(1)}_\Phi(t) = m \int dk~{d n^{(1)}_\Phi \over dk}(k,t)
\label{phi3}
\end{equation}
where
\begin{equation}
{d n^{(1)}_\Phi \over dk}(k,t) =
{C B^2 \over 4 \pi^2 k} |I(k)|^2
\left({t_1 \over t}\right)^{3 \over 2}
\label{phi4}  
\end{equation}
is the number density of excited quanta per unit comoving 
wavevector magnitude.

\subsection{Resonant excitation for $k<k_c$}

Modes with $k<k_c$ are resonantly excited at time $t_k$
given in Eq.~(\ref{tke}).  The resonance proper appears only 
in $I_+(k)$. We evaluate that integral using the saddle 
point approximation.  Near time $t_k$ 
\begin{eqnarray}
h_+(k,t) &=& h_+(k,t_k) + 
{1 \over 2} {d^2 h_+ \over d t^2}(t_k)(t - t_k)^2 + ...
\nonumber\\
&=& h_+(k,t_k) + {1 \over 4} \left({d m \over dt}(k,t_k) 
+ {m(t_k) \over 2 t_k}\right)(t - t_k)^2 + ...
\label{Taylor}
\end{eqnarray}
Neglecting the higher order terms in Eq.~(\ref{Taylor})  
and the slow variation of $g_+(k,t)$ and $\omega(k,t)$ 
inside the integral for $I_+(k)$ in Eq.~(\ref{Ike}), we obtain 
\begin{equation}
I_r(k) \simeq {2 \sqrt{\pi} g_+(k,t_k) \over 
\sqrt{\omega(k,t_k)({d m \over dt}(t_k) + {m(t_k) \over 2 t_k})}}
e^{i(h_+(k,t_k) + {\pi \over 4} + \delta)}
\label{ak}
\end{equation}
for the contribution to $I(k)$ from the resonance proper.
The resonance proper occurs over an interval of time
\begin{equation}
\Delta t_k = {1 \over \sqrt{{d^2 h_+ \over dt^2}(k,t_k)}} 
= {2 \over \sqrt{5}} \sqrt{t_2 \over m} 
\left({k_c \over k}\right)^{1 \over 5}~~\ .
\label{dtk}
\end{equation}
For $k < k_c$, we have
\begin{equation}
g_+(k,t_k) = - {1 \over 4 \sqrt{t_1 m}}
[12 m^2 \left({k \over k_c}\right)^{6 \over 5}
+ {9 \over 2 t_2^2} \left({k_c \over k}\right)^{6 \over 5}]
\label{gke}  
\end{equation}
and
\begin{equation}
\omega(k,t_k) ({dm \over dt}(t_k) + {m(t_k) \over 2 t_k})
= {5 m^2 \over t_2} \left({k \over k_c}\right)^{6 \over 5}~~\ .
\label{fke}
\end{equation}
Hence the contribution of the resonance proper to the 
spectrum of excited axions is
\begin{equation}   
{d n^{(1)}_r \over dk}(k,t)  \simeq {9 C B^2 \over 5 \pi t_1 k}
m t_2 \left({k \over k_c}\right)^{6 \over 5} \left(1 +
{3 \over 8 (t_2 m)^2}\left({k_c \over k}\right)^{12 \over 5}\right)^2
\left({t_1 \over t}\right)^{3 \over 2}~~\ .
\label{dke}
\end{equation}
For $k >> {1 \over t_1}$, when our calculation is valid,
the second term in the last parenthesis in Eq.~(\ref{dke})
is always much less than 1, because
\begin{equation}
{3 \over 8 (t_2 m)^2}(k_c t_1)^{12 \over 5} \simeq 1.4
\label{07}
\end{equation}
for all values of $m$.  We neglect henceforth the second
term in the last parenthesis in Eq.~(\ref{dke}).  We have then
\begin{equation}
n^{(1)}_r|_{k<k_c}(t) \equiv \int^{k_c} dk 
{d n^{(1)}_r \over dk}(k,t) \simeq
 {3 C B^2 \over 2 \pi t_1} m t_2 \left({t_1 \over t}\right)^{3 \over 2}
\label{idke}
\end{equation}
for the density of axions with $k<k_c$ that are excited by the 
resonance proper.  As a fraction of the density of axions in the 
zero mode [$n^{(0)} = {B^2 \over 2 t_1}$; see Eq.~(\ref{enden0})] 
this is
\begin{equation}
{n^{(1)}_r|_{k<k_c} \over n^{(0)}}
\simeq {3C \over \pi} m t_2
\simeq 0.7 \cdot 10^{-4} \left({m \over \mu{\rm eV}}\right)~~\ .
\label{fdke}
\end{equation}
Fig. 3 shows a log-log plot of the spectrum of excited
axions.  The axions excited by the resonance proper 
dominate in the region $k_* < k < k_c$.  The region 
$t_1^{-1} < k < k_*$ is dominated by axions excited 
by the near resonance effect, as discussed in the 
next subsection.  For $k > k_c$ the density of 
excited axions is very small, as discussed in 
Subsection IV.D. 

\subsection{Near resonance excitation for $k < k_c$}

Fig. 4 shows that the frequencies $\omega_\pm(k,t)$
approach $\omega(k,t)$ after the resonance proper has 
occurred.  This produces a near resonance excitation 
of the axion field whose strength is estimated in this 
subsection.  We will see that the near resonance effect 
is strongest at times shortly before $t_2$ when the axion 
mass is still turning on and $\omega_\pm(k,t)$ are both
closest to $\omega(k,t)$. Let us assume $k$ is much less 
than $k_c$ so that the time $t_k$ of the resonance 
proper occurs well before $t_2$.  We are not so 
interested in $k$ near $k_c$ because there is 
little time for the near resonance effect to occur
in that case.  The resonance proper dominates for 
$k$ near $k_c$.  We choose a time $\tilde{t}$ long 
after $t_k$ but long before $t_2$.  The precise value 
of $\tilde{t}$ will be irrelevant. The contribution of 
the near resonance effect to the integrals $I_\pm(k)$ is 
\begin{equation}
I_{\pm, nr}(k) = \int_{\tilde{t}}^\infty dt 
{g_\pm(k,t) \over \sqrt{\omega(k,t)}}
e^{i(h_\pm(k,t) + \delta)}~~\ .
\label{Ipmnr}
\end{equation}
We neglect the last two terms in the brackets 
on the RHS of Eq.~(\ref{gkt}), as we did in the 
previous subsection, because they are much smaller 
than the first term when $k >> 1/t_1$.  We have then
\begin{equation} 
g_\pm(k,t) = {3 \over 8 \sqrt{t_1 m(t)}} {dm^2 \over dT}(t) T(t)
= - {3 \over \sqrt{t_1}} m(t)^{3 \over 2} \Theta(t_2 -t)~~\ , 
\label{gnr}
\end{equation}
where $\Theta(x)$ is the Heaviside step function.
We approximate $\omega(k,t) \simeq m(t)$ since 
$k^2 {t_1 \over t}<< m(t)^2$ when $t$ approaches $t_2$.
We have then 
\begin{equation}
h_\pm(k,t) \simeq h_\pm(k,\tilde{t}) 
\pm {2 k \over \sqrt{3}}(\sqrt{t_1 t} - \sqrt{t_1 \tilde{t}}~)
\label{hnr}
\end{equation}
and 
\begin{eqnarray}
I_{\pm, nr}(k) &\simeq& - {3 \over \sqrt{t_1}}
\int_{\tilde{t}}^{t_2} dt~ m(t) 
e^{i(\tilde{\delta}_\pm(k) \pm {2k \over \sqrt{3}} \sqrt{t_1 t})}
\nonumber\\
&\simeq& \pm i {3 \sqrt{3} m \over t_1 k} \sqrt{t_2} 
e^{\pm i {2k \over \sqrt{3}}\sqrt{t_1 t_2} + i \tilde{\delta}_\pm(k)} 
[1 + {\cal O}\left(\left({\tilde{t} \over t_2}\right)^{5 \over 2},
{1 \over k \sqrt{t_1 t_2}}\right)]~~\ , 
\label{Inr2}
\end{eqnarray}
where
\begin{equation}
\tilde{\delta}_\pm(k) = \delta + h_\pm(k,\tilde{t}) 
\mp {2 k \over \sqrt{3}} \sqrt{t_1 \tilde{t}}~~\ .
\label{phases}
\end{equation}
We add $I_+(k)$ and $I_-(k)$ in quadrature in Eq.~(\ref{phi4}), 
i.e. we set $|I(k)|^2 = |I_+(k)|^2 + |I_-(k)|^2$, because their 
relative phases vary rapidly with $k$.  This yields
\begin{eqnarray}
{d n^{(1)}_{nr} \over dk}(k,t) &\simeq&
{C B^2 \over 2 \pi^2 k} 
\left({3 \sqrt{3} m \sqrt{t_2} \over t_1 k}\right)^2
\left({t_1 \over t}\right)^{3 \over 2}
\nonumber\\
&\simeq& {9 C B^2 \over 2 \pi^2 t_1} {1 \over k} 
\left({k_c \over k}\right)^2 \left({t_1 \over t}\right)^{3 \over 2}~~\ .
\label{nrfin}
\end{eqnarray}
The density of axions produced by the near resonance effect is thus
\begin{equation}
n^{(1)}_{\rm nr}(t) = \int_{k_{IR}}^{k_c} dk~
{d n^{(1)}_{nr} \over dk}(k,t) 
\simeq {9 \over 4 \pi^2} {C B^2 \over t_1}
\left({k_c \over k_{IR}}\right)^2 \left({t_1 \over t}\right)^{3 \over 2}
\label{nrtot}
\end{equation}
where $k_{IR}$ is an infrared cutoff, of order $1/t_1$, below which 
our calculation is unreliable.  We solved Eq.~(\ref{Beom}) numerically
to validate the analytical estimates.  The numerical work does not make 
the $k t_1 >> 1$ approximations made in the analytical work.  The 
numerical results, shown as grey dots in Fig. 3, are in good qualitative 
agreement with the analytical results even for $k$ near $t_1$, implying 
that $k_{IR} \simeq 1/t_1$. 

As a fraction of the axions in the zero mode, the number of axions 
excited by the near resonance effect is 
\begin{equation}
{n^{(1)}_{nr} \over n^{(0)}} \simeq 
{9 C \over 2 \pi^2} \left({k_c \over k_{IR}}\right)^2
\simeq 0.06 \left({1 \over t_1 k_{IR}}\right)^2
\left({m \over \mu{\rm eV}}\right)^{5 \over 3}~~\ .
\label{nr0}
\end{equation}
Since $k_{IR}$ is of order $t_1$, the fraction of axions 
that get excited is of order one if the axion mass is larger 
than a few $\mu$eV.  The near resonance effect dominates 
for $k<k_*$ with
\begin{equation}
k_* = k_c \left({5 \over 2 \pi m t_2}\right)^{5 \over 16}
\simeq {k_c \over 22} \left({\mu{\rm eV} \over m}\right)^{5 \over 16}
\label{kstar}
\end{equation}
whereas the resonance proper dominates for $k>k_*$.

\subsection{Resonant excitation for $k > k_c$}

We estimate here the density of axions that 
are excited after $t_2$, i.e. for $k>k_c$.  
The time at which the resonance proper
occurs is given in Eq.~(\ref{tkl}).  The same 
expression for $I_+(k)$, Eq.~(\ref{ak}), applies 
here but with 
\begin{equation}
g_+(k,t_k) = - {9 \over 8 t_2^2 \sqrt{t_1 m}} 
\left({k_c \over k}\right)^4
\label{gkl}
\end{equation}
and 
\begin{equation}
\omega(t_k) ({dm \over dt}(t_k) + {m(t_k) \over 2 t_k})
= \left({k_c \over k}\right)^2 {m^2 \over t_2}~~\ .
\label{denkl}
\end{equation}
Hence the axions excited by the resonance proper 
have spectral density 
\begin{equation}
{d n^{(1)}_r \over dk}(k,t) \simeq 
{81 C B^2 \over 64 \pi t_1 k} {1 \over (t_2 m)^3}
\left({k_c \over k}\right)^6\left({t_1 \over t}\right)^{3 \over 2}
\label{bk}
\end{equation}
for $k>k_c$, and density
\begin{equation}
n^{(1)}_r|_{k>k_c}(t) \equiv 
\int_{k_c}^\infty {d n^{(1)}_r \over dk}(k,t)
\simeq {27 C B^2 \over 128 \pi t_1} {1 \over (t_2 m)^3}
\left({t_1 \over t}\right)^{3 \over 2}~~\ .
\label{inbk}
\end{equation}
As a fraction of the axions in the zero mode this is
\begin{equation}
{n^{(1)}_r|_{k>k_c} \over n^{(0)}} \simeq
{27 C \over 64 \pi} {1 \over (t_2 m)^3}
= 2 \cdot 10^{-22} \left({\mu {\rm eV} \over m}\right)^3
\label{bkf}
\end{equation}
which is always very small.  For most plausible values of 
$H_I$, it is much smaller than the contribution from 
the quantum mechanical fluctuations in the axion field, 
discussed in Section III.  The reason why resonant excitation 
is such a small effect for $k>k_c$ is that the axion mass is 
constant after $t_2$ and hence the term $T {dm^2 \over dT}$ 
is absent from $g_+(k,t)$.  For $k > k_c$ the contribution 
from the near resonance effect is found to be even smaller 
than the small contribution from the resonance proper.

\section{Summary}

We derived the spectrum of cold axions after the QCD phase 
transition has been completed for wavevectors $k$ at time $t_1$
much larger than $1/t_1$. $t_1$ was defined in Eq.~(\ref{time1}).
Two kinds of axion field perturbations contribute to the spectrum 
in case inflation occurs after the PQ phase transition.  First, 
there are the perturbations that originate as quantum mechanical 
fluctuations in the axion field during the inflationary phase.  
Those were treated in Section III.  Their contribution to the 
cold axion energy density is very small when the isocurvature 
constraints from the CMB anisotropy observations are taken into 
account.  Second, there are the fluctuations induced in the 
axion field by the adiabatic perturbations in the primordial 
plasma.  These were treated in Section IV.  We show that the 
adiabatic perturbations in the plasma resonantly pump low 
momentum axions into modes of wavevector magnitude up to of 
order $\sqrt{3} m$ where $m$ is the zero temperature axion 
mass.  The process conserves axion number and transfers energy 
from the primordial plasma to the axion fluid.  The spectrum 
of axions that are resonantly excited is given in Eqs.~(\ref{dke}), 
(\ref{nrfin}) and (\ref{bk}), and illustrated in Fig. 3.  The 
fraction of cold axions that get excited is of order one if 
the axion mass is larger than a few $\mu$eV.

The process of axion field excitation occurs in qualitatively 
the same way whether inflation occurs before or after the PQ 
phase transition since the initial cold axion momenta are in 
either case of order $1/t_1$ or less, much smaller than the 
momenta to which they get resonantly excited.

Resonant excitation of the axion fluid during the QCD phase 
transition does not change the status of the axion as a cold 
dark matter candidate since the excited axions are non-relativistic 
again long before the onset of large scale structure formation.   
However, the excitation spectrum may have implications for 
observation.  In future work, we intend to study its implications 
for the formation of axion mini-clusters \cite{KT} and for the 
cosmic time at which cold axion dark matter first Bose-Einstein 
condenses by gravitational self-interactions \cite{CABEC}.

\begin{acknowledgments}

PS thanks Richard Woodard for useful discussions.  This 
work was supported in part by the U.S. Department of Energy 
under grant DE-SC0022148 at the University of Florida. 

\end{acknowledgments}


\newpage

\begin{thebibliography}{bib}

\bibitem{PQ}
R. D. Peccei and H. Quinn, Phys. Rev. Lett. 38 (1977) 1440
and Phys.Rev. D16 (1977) 1791.

\bibitem{WW}
S. Weinberg, Phys. Rev. Lett. 40 (1978) 223; 
F. Wilczek, Phys. Rev. Lett. 40 (1978) 279.

\bibitem{inv}
J. Kim, Phys. Rev. Lett. 43 (1979) 103; M. A. Shifman,
A. I. Vainshtein and V. I. Zakharov, Nucl. Phys. B166
(1980) 493; M. Dine, W. Fischler and M. Srednicki, Phys. Lett. 
B104 (1981) 199; A. Zhitnitskii, Sov. J. Nucl. 31 (1980) 260.

\bibitem{axdm}
J. Preskill, M. Wise and F. Wilczek, Phys. Lett. B120 (1983) 127;
L. Abbott and P. Sikivie, Phys. Lett. B120 (1983) 133;
M. Dine and W. Fischler, Phys. Lett. B120 (1983) 137.

\bibitem{axrev}
Reviews of axion cosmology include:  
P. Sikivie, Lect. Notes Phys. 741 (2005) 083513;
D.J.E. Marsh, Phys. Rep. 643 (2016) 1.

\bibitem{GPY}
D.J. Gross, R.D. Pisarski and L.G. Yaffe, Rev. Mod. Phys. 53 (1981) 43.

\bibitem{Bors}
Sz. Borsanyi et al., Nature 539 (2016) 69.

\bibitem{Bonati}
C. Bonati et al., EPJ Web Conf. 137 (2017) 08004.

\bibitem{infl}
A.A. Starobinsky, Phys. Lett. B91 (1980) 99; 
D. Kazanas, Ap. J. 241 (1980) L59;
K. Sato, MNRAS 195 (1981) 467;

\bibitem{infl2}
A. Guth, Phys. Rev. D23 (1981) 347;
A.D. Linde, Phys. Lett. B108 (1982) 389;
A. Albrecht and P. Steinhardt, Phys. Rev. Lett. 48 (1982) 1220. 

\bibitem{isoc}
M. Axenides, R.H. Brandenberger and M.S. Turner, 
Phys. Lett. 126 (1983) 178;
P.J. Steinhardt and M.S. Turner, Phys. Lett. B129 (1983) 51;
A.D. Linde, Phys. Lett. B158 (1985) 375;
D. Seckel and M.S. Turner, Phys. Rev. D32 (1985) 3178;
D.H. Lyth, Phys. Lett. B236 (1990) 408;
M.S. Turner and F. Wilczek, Phys. Rev. Lett. 66 (1991) 5.

\bibitem{adiab}
V.F. Mukhanov and G.V. Chibisov, JETP Lett. 33 (1981) 532;
S.W. Hawking, Phys. Lett. B115 (1982) 295;
A. Guth and S.-Y. Pi, Phys. Rev. Lett. 49 (1982) 1110;
J.M. Bardeen, P.J. Steinhardt and M.S. Turner, Phys. Rev. D28 (1983) 679.

\bibitem{Husdal}
L. Husdal, Galaxies 4 (2016) 78.

\bibitem{Dodelson}
S. Dodelson, {\it Modern Cosmology}, Academic Press, Elsevier 2003.

\bibitem{KT}
E.W. Kolb and I.I. Tkachev, Phys. Rev. Lett. 71 (1993) 3051.

\bibitem{CABEC}
P. Sikivie and Q. Yang, Phys. Rev. Lett. 103 (2009) 111301;
O. Erken, P. Sikivie, H. Tam and Q. Yang, Phys. Rev. D85 (2012) 063520.

\end{thebibliography}
\end{document}